\documentclass[11pt]{article}

\usepackage[margin=0.9in,top=0.75in,bottom=0.8in]{geometry}
\usepackage{amsmath,amssymb,mathtools}
\usepackage{booktabs,longtable,array,multirow}
\usepackage{graphicx}
\usepackage{xcolor}
\usepackage{tikz}
\usepackage{float}
\usetikzlibrary{arrows.meta,positioning,calc}
\usepackage{float}
\usepackage{caption}
\usepackage{subcaption}
\usepackage{authblk}
\usepackage{xurl}
\usepackage[hidelinks]{hyperref}
\usepackage{microtype}
\usepackage{enumitem}
\usepackage{listings}
\usepackage{cite}
\usepackage{titlesec}

\definecolor{draftgray}{gray}{0.94}
\definecolor{darkblue}{RGB}{25,55,100}
\definecolor{darkred}{RGB}{125,35,35}

\setlist{nosep}
\titleformat{\paragraph}[runin]{\normalfont\normalsize\bfseries}{}{0pt}{}[.]
\titlespacing*{\paragraph}{0pt}{0.7\baselineskip}{0.6em}

\newcommand{\PF}{\mathrm{PF}}
\newcommand{\RoM}{\mathrm{RoM}}
\newcommand{\LRR}{\mathrm{LRR}}
\newcommand{\Var}{\operatorname{Var}}

\newcommand{\data}{\mathcal{D}}

\title{\textbf{Fluoride Varnish in a Changing Preventive Environment:
A Reanalysis and Temporal Meta-Regression of Historical and Contemporary Trials}}

\author{\Large Sara Antonijevic}
\author{\Large Danielle Sitalo}
\author{\Large Brani Vidakovic}
\affil{Department of Statistics, Texas A\&M University\\
College Station, TX 77843, USA\\
Correspondence: saraant@tamu.edu}
\date{\today}

\begin{document}
\maketitle

\begin{abstract}

\textbf{Background.}
Historical trials of fluoride varnish reported substantial reductions in
dental caries, whereas many contemporary trials are conducted against more
intensive background prevention. The relevant question is therefore not only whether
fluoride varnish is effective, but whether its \emph{incremental} contribution
remains as large as in earlier trials.

\textbf{Methods.}
For positive continuous caries outcomes, the principal effect measure is the
log ratio of treatment and control means,
$y_i=\log(\bar X_{Ti}/\bar X_{Ci})$, with results reported on the
prevented-fraction scale, $\PF_+=1-\exp(y)$. A nine-trial historical dataset is
reanalyzed using REML with Hartung--Knapp inference, Bayesian hierarchical
models, aggregate arm-level modeling, heavy-tailed random effects, prediction,
and influence analyses. Historical completeness is assessed on the
Cochrane-compatible direct prevented-fraction scale using all 13
permanent-surface trials in the corresponding synthesis. Temporal evolution is then examined in a 14-study continuous-outcome dataset comprising the nine
historical trials and five contemporary comparisons, including Wang with an
approximately cluster-adjusted sampling variance. Alternative temporal
specifications are compared by maximum-likelihood AICc. Binary caries-incidence
outcomes and study-level moderators are analyzed separately as supporting
analyses.

\textbf{Results.}
The historical nine-trial REML analysis gives a pooled prevented fraction of
$39.7\%$ (Hartung--Knapp 95\% CI $18.0\%$ to $55.6\%$), with a substantially
wider 95\% prediction interval of $-26.5\%$ to $71.2\%$. On the direct
prevented-fraction scale, expanding the historical evidence set from nine to
13 Cochrane trials changes the pooled estimate only from $0.439$ to $0.435$.
In the 14-study temporal analysis, the REML--Hartung--Knapp calendar coefficient
is $0.0986$ log-ratio units per decade (95\% CI $-0.0445$ to $0.2417$;
residual $\widehat{\tau}=0.279$). The corresponding fitted prevented fraction
decreases from $34.8\%$ in 2005 to $20.6\%$ in 2025. A Bayesian calendar-time
model gives a posterior median slope of $0.091$ (95\% CrI $-0.036$ to $0.231$)
and $\Pr(\beta_1>0\mid\data)=0.924$. The intercept-only model has the smallest
AICc, although the linear-time and historical-versus-contemporary models are
both within two AICc units. Adjustment for protocol-specified background
prevention reduces the posterior median time slope from $0.091$ to $0.057$ per
decade, although substantial uncertainty remains.

\textbf{Conclusions.}
The historical average preventive effect is stable across several modeling
choices, whereas prediction to a new setting is considerably less certain.
The contemporary evidence is compatible with attenuation of the incremental
varnish effect over calendar time, but does not statistically require a
temporal moderator. The observed pattern is consistent with a changing treatment contrast in
which fluoride varnish is added to more intensive background prevention and
should not, by itself, be interpreted as evidence that the biological action
of fluoride varnish has weakened.

\end{abstract}


\section{Introduction}
\label{sec:introduction}

Fluoride varnish has long been used to prevent dental caries in children and
adolescents. Its practical appeal is clear: application is intermittent, can be
delivered in clinical, school, or community settings, and does not depend on
daily adherence. The resulting trial literature, however, spans several decades
during which baseline caries burden, background fluoride exposure, preventive
care, diagnostic practice, dentition, and trial methodology have changed
substantially.

The historical literature generally suggested a large average preventive
effect. The 2013 Cochrane review reported a pooled prevented fraction of
approximately $43\%$ for permanent tooth surfaces, with substantial
heterogeneity~\cite{Marinho2013}. Later evidence has been less uniform. A
systematic review focused on preschool children found more modest effects at
the individual level~\cite{Sousa2019}, and a recent overview of systematic
reviews emphasized variation in both estimated effectiveness and methodological
quality. In particular, newer evidence from children already exposed to routine
fluoride has tended to show a smaller incremental benefit than some of the
older trials~\cite{Couto2026}.

These observations motivate a distinction between the \emph{historical average
treatment effect} and the \emph{contemporary incremental effect}. An older
trial could compare varnish with relatively limited background prevention,
whereas a contemporary trial may compare varnish added to fluoride toothpaste,
oral-health education, supervised brushing, fluoridated water, or an organized
preventive program with that same background care alone. A smaller contemporary
treatment contrast is therefore compatible with preserved biological activity
of fluoride varnish if the control condition has become more effective.

A ratio-based estimand is natural for this question. If $\bar X_{Ti}$ and
$\bar X_{Ci}$ are positive treatment and control means in study $i$, we analyze
the log ratio of means
$$
y_i=\LRR_i
=
\log\left(\frac{\bar X_{Ti}}{\bar X_{Ci}}\right),
$$
and report results on the clinically interpretable prevented-fraction scale
$$
\PF_{+,i}=1-\exp(y_i).
$$
Negative $y_i$ favors fluoride varnish, while $y_i=0$ is the null. Equivalently,
positive values of $\PF_{+,i}$ favor varnish. This parameterization respects
the multiplicative structure of the effect and avoids fitting a symmetric
Gaussian likelihood directly to the nonlinear, upper-bounded
prevented-fraction scale~\cite{Friedrich2011,Bakbergenuly2020,Lajeunesse2015}.

Calendar time provides a simple first assessment of whether the treatment
contrast has changed. A positive calendar coefficient on the log-ratio scale
moves the treatment-to-control ratio toward one and therefore moves $\PF_+$
toward zero over the observed calendar range. Calendar year is not interpreted
as a causal exposure; it is a study-level marker for multiple secular changes,
including baseline risk, comparator intensity, dentition, follow-up, trial
design, and background prevention.

The analysis proceeds in four stages. First, we establish a historical
reference effect using the nine trials for which arm-level means, standard
deviations, and sample sizes are available on a sufficiently common scale for
the preferred log-ratio analysis. Second, we assess historical completeness
using the 13 permanent-tooth-surface trials represented in the corresponding
Cochrane synthesis. Third, we extend the continuous-outcome analysis to a 14-study primary
temporal dataset comprising the nine historical trials and five contemporary
comparisons with sufficiently compatible continuous outcomes and usable
sampling variances. Two additional contemporary studies with incompletely
identified design-adjusted variances are retained for sensitivity analysis. Finally, we examine
binary caries-incidence outcomes and parsimonious study-level moderators as
supporting analyses rather than forcing heterogeneous endpoints into a single
pooled estimand.

This is not a new systematic review, and the temporal extension should not be
read as a definitive estimate for all contemporary fluoride-varnish trials.
Its purpose is narrower: to ask whether the available historical and
contemporary evidence is compatible with a common incremental treatment effect
and, if not, whether measurable features of the preventive environment help
explain the observed difference.

The remainder of the paper is organized as follows. Section~2 develops the
statistical framework, including the log-ratio estimand, frequentist and
Bayesian random-effects formulations, robustness analyses, and the
calendar-time meta-regression. Section~3 establishes the historical reference
effect and examines its robustness to alternative modeling assumptions and to
expansion from the nine-study core to the complete 13-trial Cochrane
permanent-surface evidence set. Section~4 extends the analysis to contemporary
continuous and binary outcomes and evaluates the evolution of the incremental
preventive effect over calendar time. Section~5 examines potential explanations
for the observed temporal pattern, with particular attention to strengthening
background prevention and study-level moderator analyses. Sections~6 and~7
provide the discussion and conclusions, respectively.

\section{Statistical Framework}
\label{sec:methods}

The principal analyses are based on a normal--normal hierarchical
random-effects model for study-specific log treatment-to-control ratios. The
same underlying model is examined from both frequentist and Bayesian
perspectives. REML with Hartung--Knapp inference provides the principal
frequentist analysis, while Bayesian hierarchical inference is used to
quantify heterogeneity, posterior prediction, and sensitivity to prior
specification.

We first define the log-ratio effect measure and hierarchical model, then
describe frequentist and Bayesian inference and posterior predictive
quantities. Finally, we assess sensitivity to the heterogeneity prior, the
first-stage log-ratio approximation, the assumed distribution of study-specific
effects, and the construction of the contemporary temporal analysis.

 \subsection{Log-ratio effect and hierarchical random-effects model}
\label{sec:effect-random}

For study $i$, let $\bar X_{Ti}$ and $\bar X_{Ci}$ denote the reported
mean caries outcomes in the fluoride-varnish and control groups,
respectively. The principal relative effect is the ratio of means,
$$
\RoM_i=\frac{\bar X_{Ti}}{\bar X_{Ci}},
$$
and inference is performed on the log-ratio scale,
$$
y_i=\LRR_i
=
\log\left(
\frac{\bar X_{Ti}}{\bar X_{Ci}}
\right).
$$
The null value is $y_i=0$, and negative values favor fluoride varnish.
For clinical interpretation, results are transformed to the positive
prevented-fraction scale,
$$
\PF_{+,i}
=
1-\RoM_i
=
1-\exp(y_i).
$$
Thus, for example, $\PF_+=0.40$ corresponds to a treatment mean that is
$40\%$ lower than the control mean.

The log ratio of means provides a natural multiplicative effect measure for
positive continuous outcomes and avoids imposing a symmetric Gaussian model
directly on a bounded prevented fraction~\cite{Friedrich2011,Bakbergenuly2020}.
Using a first-order delta approximation, the sampling variance is
$$
s_i^2
=
\Var(y_i)
\approx
\frac{s_{Ti}^2}{n_{Ti}\bar X_{Ti}^{\,2}}
+
\frac{s_{Ci}^2}{n_{Ci}\bar X_{Ci}^{\,2}},
$$
where $s_{Ti}$ and $s_{Ci}$ are the reported arm-specific standard deviations.
As with other response-ratio estimators, this approximation can become less
accurate when an arm mean is small relative to its standard deviation.
Bias-corrected response-ratio calculations are therefore retained as a
supplementary sensitivity analysis rather than used to define a second primary
estimand~\cite{Bakbergenuly2020,Lajeunesse2015}.

The study-specific observed effects are modeled using the conventional
normal--normal random-effects formulation,
$$
y_i\mid\theta_i
\sim
N(\theta_i,s_i^2),
\qquad
\theta_i\mid\mu,\tau
\sim
N(\mu,\tau^2),
\qquad i=1,\ldots,k.
$$
Here $\mu$ is the mean log treatment-to-control ratio across the population of
studies and $\tau$ is the between-study standard deviation. The corresponding
average prevented fraction is
$$
\PF_{+,\mathrm{pool}}
=
1-\exp(\mu).
$$

For the frequentist benchmark, $\tau^2$ is estimated by restricted maximum
likelihood (REML). Uncertainty for $\mu$ is based on the Hartung--Knapp
small-sample procedure rather than a normal Wald approximation. This is
particularly relevant in the present setting because the historical analysis
contains relatively few studies and the uncertainty in between-study
heterogeneity is consequently non-negligible~\cite{RoverKnappFriede2015}.
The modified Hartung--Knapp safeguard is also examined; in the present data it
does not alter the interval because the Hartung--Knapp scale factor exceeds
one.

In addition to an interval for the average effect, we report a prediction
interval for the true effect in a new comparable study. The distinction is
substantive rather than merely technical. A confidence interval for $\mu$
quantifies uncertainty about the mean effect across studies, whereas a
prediction interval incorporates both uncertainty in that mean and genuine
between-study variation~\cite{IntHout2016}. All intervals are constructed on
the log-ratio scale and then transformed to the prevented-fraction scale for
presentation.

\subsection{Bayesian specification and prediction}
\label{sec:bayes-nnhm}

The Bayesian analysis uses the same normal--normal likelihood,
$$
y_i\mid\theta_i
\sim
N(\theta_i,s_i^2),
\qquad
\theta_i\mid\mu,\tau
\sim
N(\mu,\tau^2),
$$
with the primary prior specification
$$
\mu\sim N(0,1^2),
\qquad
\tau\sim\mathrm{Half\mbox{-}Normal}(0,0.5^2).
$$

The prior on $\mu$ is intentionally broad on the multiplicative scale. A 95\%
prior interval of approximately $[-1.96,1.96]$ corresponds to
treatment-to-control mean ratios ranging from about $0.14$ to $7.1$, far wider
than effects considered plausible in the present application.

The prior for $\tau$ is more consequential because heterogeneity is weakly
identified when the number of studies is small. We therefore calibrate this
prior on the effect scale rather than describing it simply as ``vague'' or
``informative.'' If the true study effects are normally distributed with
standard deviation $\tau$, approximately 95\% lie within
$\mu\pm1.96\tau$. Thus $\tau=0.25$ corresponds to multiplicative departures
from the pooled ratio by factors of approximately

$$
\exp(1.96\times0.25)=1.63,
$$

whereas $\tau=0.50$ corresponds to a factor of approximately

$$
\exp(1.96\times0.50)=2.66.
$$
The primary half-normal prior therefore permits substantial heterogeneity while
progressively downweighting extremely large values. Such weakly informative,
effect-scale-calibrated priors are particularly useful when the likelihood
contains limited information about a variance component~\cite{Gelman2006,RoverEtAl2021,Lilienthal2024}.

The posterior distribution of the pooled effect is summarized through
$$
\PF_{+,\mathrm{pool}}
=
1-\exp(\mu).
$$
More importantly for transportability, inference is also based on the posterior
predictive distribution for a new true study effect,
$$
\theta_{\mathrm{new}}\mid\mu,\tau
\sim
N(\mu,\tau^2),
\qquad
\PF_{+,\mathrm{new}}
=
1-\exp(\theta_{\mathrm{new}}).
$$
In addition to posterior medians and credible or predictive intervals, we
report quantities such as
$$
\Pr(\PF_{+,\mathrm{new}}>0\mid\mathcal D)
$$
and
$$
\Pr(\PF_{+,\mathrm{new}}>0.20\mid\mathcal D),
$$
which describe respectively the posterior probability of any preventive effect
and of at least a $20\%$ reduction in a new comparable study.

\subsection{Sensitivity and robustness}
\label{sec:sensitivity}

The primary model makes two assumptions that deserve particular scrutiny in a
small meta-analysis: the prior assigned to between-study heterogeneity and the
use of a Gaussian first-stage log-ratio likelihood with normally distributed
random effects. We therefore use a small number of targeted sensitivity
analyses rather than a large catalogue of alternative models.

\paragraph{Sensitivity to the heterogeneity prior.}
The primary prior
$\tau\sim\mathrm{Half\mbox{-}Normal}(0,0.5^2)$ is compared with a more
concentrated half-normal prior,
$$
\tau\sim\mathrm{Half\mbox{-}Normal}(0,0.25^2),
$$
a heavier-tailed prior,
$$
\tau\sim\mathrm{Half}\mbox{-}t_4(0,0.5),
$$
and, for continuity with the earlier analysis, the DuMouchel prior
$$
\pi(\tau)
\propto
\frac{s_0}{(s_0+\tau)^2},
\qquad
\tau>0,
$$
where $s_0$ is the prespecified reference within-study standard-error scale
defined in the Supplementary Material and analysis code~\cite{DuMouchel1994}.
The purpose is not to identify a single uniquely correct heterogeneity prior,
but to determine which substantive conclusions depend materially on reasonable
changes in prior concentration and tail behavior. Particular attention is paid
to posterior prediction, because predictive tails depend directly on $\tau$
and are expected to be more prior-sensitive than the pooled center.

\paragraph{Aggregate arm-level model}
The contrast-based analysis treats $y_i$ and its estimated sampling variance
$s_i^2$ as first-stage summaries. To examine whether the principal conclusions
depend on this approximation, we also model the reported arm means directly.
Let $m_i>0$ denote the latent control mean in study $i$ and $\theta_i$ the
study-specific log treatment-to-control mean ratio. The aggregate arm-level
model is
$$
\bar X_{Ci}\mid m_i
\sim
N\left(
m_i,\frac{s_{Ci}^2}{n_{Ci}}
\right),
$$
$$
\bar X_{Ti}\mid m_i,\theta_i
\sim
N\left(
m_i\exp(\theta_i),
\frac{s_{Ti}^2}{n_{Ti}}
\right),
$$
with
$$
\log m_i=\eta_i,
\qquad
\eta_i\sim N(0,2.5^2),
$$
and
$$
\theta_i\mid\mu,\tau
\sim
N(\mu,\tau^2).
$$
The priors for $\mu$ and $\tau$ are the same as in the primary Bayesian
analysis. This formulation generates the treatment-to-control ratio directly
through $m_i\exp(\theta_i)$ and allows baseline caries burden to vary freely
across studies. It therefore provides a useful check on whether the main result
is an artifact of the first-stage delta approximation.

This remains an aggregate-data model. The reported standard deviations are
treated as fixed, and unavailable features such as zero inflation,
within-person dependence, individual-level covariates, and cluster effects
cannot be reconstructed from published arm summaries. The model is therefore
used as a robustness analysis rather than as a substitute for individual-level
data.

\paragraph{Heavy-tailed random effects}
Finally, we examine sensitivity to the normality assumption for the
distribution of the true study effects. We replace the Gaussian random effect
by

\begin{eqnarray*}
\theta_i &=& \mu+\tau z_i,\\
z_i &\sim& t_4\left(0,\sqrt{\frac{4-2}{4}}\right),
\end{eqnarray*}

where the Student-$t$ distribution is scaled to have variance one.
Consequently, $\tau$ retains its interpretation as the between-study standard
deviation while the model assigns greater probability to unusually large
departures from the pooled effect. The degrees of freedom are fixed at four
because the number of historical studies is too small to identify an
additional tail parameter reliably. This model is used as a robustness check,
not as an assertion that the population of study effects is literally
Student-$t$ distributed~\cite{Wang2025,Panagiotopoulou2025}.

More elaborate calculations, including the direct prevented-fraction
analysis, second-order response-ratio corrections, complete study-specific
posterior shrinkage results, leave-one-study-out analyses, detailed MCMC
diagnostics, alternative heterogeneity priors, and implementation code, are
reported in the Supplementary Material. The main text retains analyses that
materially affect interpretation of the historical reference effect, its
heterogeneity, or prediction to a new study.

\subsection{Temporal extension, variance audit, and model comparison}
\label{ssec:temporal}

The primary calendar-time analysis uses 14 continuous-outcome comparisons:
the nine historical studies and the contemporary studies Mu\~noz-Mill\'an,
Latifi-Xhemajli, McMahon, Zeng, and Wang~\cite{Munoz2018,Latifi2019,McMahon2020,Zeng2025,Wang2022}. For the first four
contemporary studies, sampling variances are calculated from the reported
arm-level summaries using the same first-order log-ratio approximation as in
the historical analysis.

Wang is cluster randomized. Its primary sampling variance is therefore
approximately adjusted using the published intraclass correlation coefficient
$\rho=0.20$ and the planning cluster size $m=40$. The corresponding design
effect is

$$
DE=1+(m-1)\rho
   =1+39(0.20)
   =8.8,
$$

and the naive sampling variance is multiplied by this factor. This is an
approximately cluster-adjusted variance rather than a published cluster-robust
standard error. Replacing $m=40$ by the observed mean completed cluster size
has negligible influence on the fitted calendar slope.

Jayasinghe and Ghasemi cannot be assigned unique design-adjusted variances from
the published summaries~\cite{Jayasinghe2022,Ghasemi2025}. Jayasinghe reports the two-year means but not
a usable SD or SE for the new-caries count; Ghasemi randomized only four
schools and does not report an outcome ICC or school-specific DMFS increments.
Their effects are therefore included only in explicit overdispersion and
design-effect sensitivity analyses.

For the primary 14-study dataset, let

$$
t_i=\frac{\mathrm{Year}_i-2000}{10},
$$

where $\mathrm{Year}_i$ denotes publication year. The calendar-time
meta-regression is

\begin{eqnarray*}
y_i\mid\theta_i &\sim& N(\theta_i,s_i^2),\\
\theta_i &=& \beta_0+\beta_1t_i+u_i,\\
u_i &\sim& N(0,\tau^2).
\end{eqnarray*}

Thus $\beta_1$ is the change in log treatment-to-control ratio per decade.
Because negative log ratios favor fluoride varnish, $\beta_1>0$ corresponds
to movement toward the null value and hence to attenuation of the incremental
preventive effect over calendar time. Frequentist estimation uses REML with
Hartung--Knapp inference.

We compare an intercept-only model, the linear calendar-time model, an
indicator contrasting historical and contemporary studies, a quadratic
calendar-time model, and a three-era model. Because restricted likelihoods
with different fixed-effect design matrices are not directly comparable,
model comparison is based on maximum-likelihood AICc. REML--Hartung--Knapp
inference is then used for estimation and uncertainty under the linear
calendar-time specification. The Bayesian calendar-time model uses

$$
\beta_0\sim N(0,1^2),\qquad
\beta_1\sim N(0,0.25^2),\qquad
\tau\sim {\rm Half\mbox{-}Normal}(0,0.5^2).
$$

\section{Historical Fluoride-Varnish Evidence}
\label{sec:historical}

The historical trials provide the reference against which the contemporary
evidence is interpreted. We begin with the nine-study dataset for which
treatment and control means, standard deviations, and sample sizes are
available on a sufficiently common scale to support the preferred log-ratio
analysis and the Bayesian models developed in Section~\ref{sec:methods}. The purpose is not to
select a preferred model from a large collection of alternatives, but to
determine whether the magnitude of the historical preventive effect is stable
under reasonable changes in inferential framework, prior specification, and
random-effects distribution.

The nine-study analysis is followed by a historical-completeness check using
the 13 permanent-tooth-surface trials represented in the corresponding
Cochrane synthesis. Because several of the four additional studies cannot be
placed cleanly into the preferred log-ratio likelihood, this extension is used
only to assess whether omission of those trials materially changes the
historical reference effect.

\subsection{Study-level effects and the historical reference effect}
\label{sec:historical-main}

Table~\ref{tab:historical-effects} summarizes the study-specific effects on the
log-ratio scale. Eight of the nine point estimates favor fluoride varnish.
Milsom is the exception, with an estimated log ratio close to zero and a
slightly negative prevented fraction. Koch and Tewari show the largest
relative reductions, but their log-ratio estimates are also among the least
precise because their treatment means are small relative to the reported
within-arm variability.

\begin{table}[ht]
\centering
\caption{Study-specific historical effects. The prevented fraction is defined as
$\PF_{+,i}=1-\exp(y_i)$, so that positive values favor fluoride varnish and
negative values favor control.}
\label{tab:historical-effects}
\small
\begin{tabular}{lrrr}
\toprule
Study & Log ratio $y_i$ & SE$(y_i)$ & $\widehat{\PF}_{+,i}$ \\
\midrule
Koch        & $-1.492$ & 0.558 & 0.775 \\
Modeer      & $-0.392$ & 0.168 & 0.324 \\
Clark       & $-0.247$ & 0.110 & 0.219 \\
Tewari      & $-1.368$ & 0.475 & 0.745 \\
Bravo       & $-0.556$ & 0.125 & 0.426 \\
Sk\"old       & $-0.851$ & 0.192 & 0.573 \\
Arruda      & $-0.516$ & 0.118 & 0.403 \\
Tagliaferro & $-0.547$ & 0.422 & 0.421 \\
Milsom      & $\phantom{-}0.047$ & 0.157 & $-0.048$ \\
\bottomrule
\end{tabular}
\end{table}

\begin{figure}[htb]
\centering
\includegraphics[width=0.8\textwidth]
{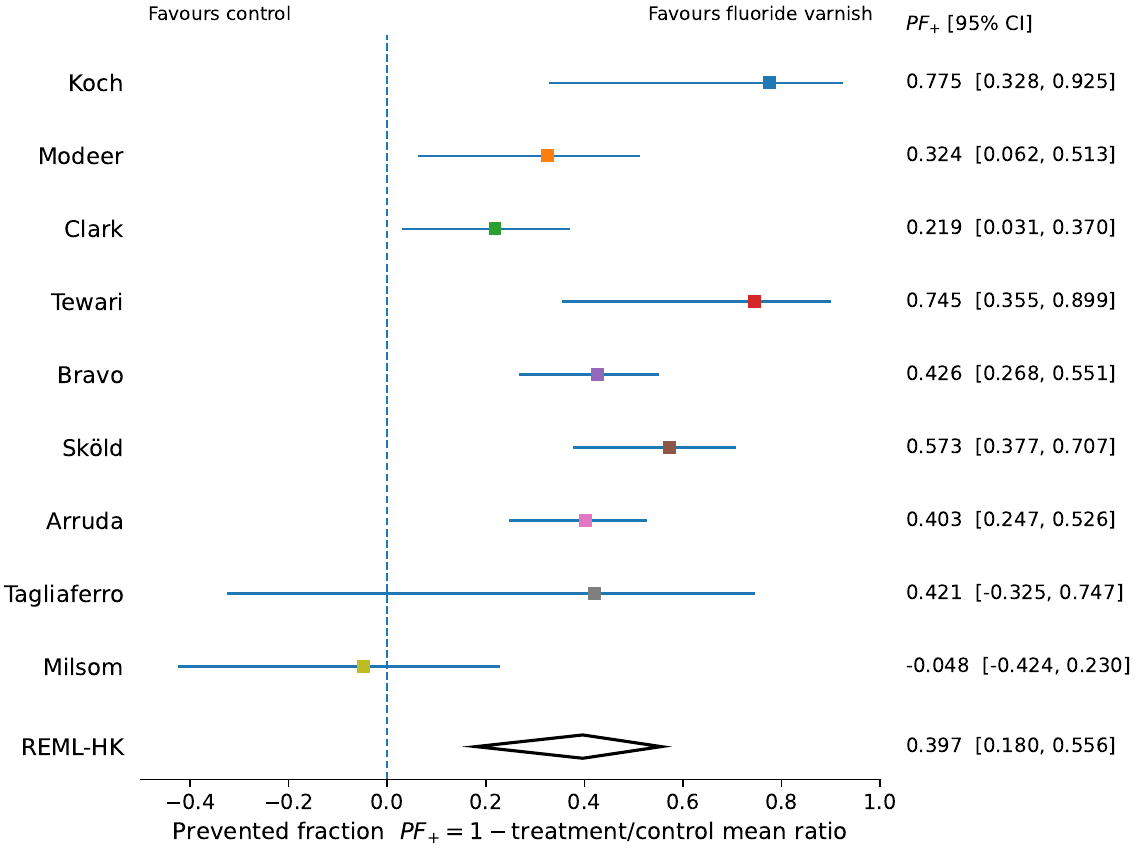}
\caption{Forest plot of study-specific prevented fractions for the nine
historical fluoride-varnish trials. Squares denote the estimated prevented
fractions, horizontal lines show the corresponding 95\% confidence intervals,
and the diamond shows the pooled REML--Hartung--Knapp estimate and its 95\%
confidence interval. The vertical dashed line at $\mathrm{PF}_{+}=0$ denotes
the null effect; positive values favor fluoride varnish.}
\label{fig:historical_forest}
\end{figure}

The same pattern is apparent in Figure~\ref{fig:historical_forest}.
Eight of the nine study-specific point estimates favor fluoride varnish,
although their precision varies considerably across trials; Milsom is the
only study with a point estimate close to the null. The pooled
REML--Hartung--Knapp estimate remains clearly on the preventive side of the
null.

The frequentist random-effects benchmark gives
$$
\widehat\mu_{\rm REML}=-0.5050,
\qquad
\widehat\tau=0.2835.
$$
Transforming the estimated mean log ratio gives
$$
1-\exp(\widehat\mu_{\rm REML})=0.3965,
$$
corresponding to an estimated historical preventive fraction of approximately
$39.7\%$. The Hartung--Knapp 95\% confidence interval is
$ 
[0.180,\;0.556].
$ 

The corresponding 95\% prediction interval for a true effect in a new
comparable study is much wider,
$ 
[-0.265,\;0.712].
$ 

Thus uncertainty about the pooled historical mean is substantially smaller than
uncertainty about the true effect in a new comparable setting. This distinction
between estimation of the pooled center and prediction will recur throughout
the Bayesian analyses.

The historical pooled center is also relatively stable under deletion of a
single trial. Across the nine leave-one-study-out REML analyses, the pooled
prevented fraction ranges from $0.357$ to $0.430$. Prediction is more
sensitive. In particular, removal of Milsom reduces $\widehat{\tau}$ from
$0.284$ to approximately $0.183$ and changes the prediction interval to
approximately $[0.030,0.659]$. Thus Milsom has appreciable influence on
estimated heterogeneity and transportability, but considerably less influence
on the pooled historical center.
The complete leave-one-study-out table is given in the Supplementary Material.

\subsection{Bayesian synthesis, sensitivity, and robustness}
\label{sec:historical-bayes}

The contrast-based Bayesian normal--normal hierarchical model was evaluated
with \texttt{bayesmeta}, which exploits the conditional Gaussian structure and
uses deterministic numerical integration over the heterogeneity parameter
rather than MCMC~\cite{Rover2020}. Under the primary prior
$\tau\sim\mathrm{Half\mbox{-}Normal}(0,0.5^2)$, the posterior median pooled
prevented fraction is
$ 
0.389,
$ 
with 95\% credible interval
$ 
[0.215,\;0.556].
$ 

The posterior median heterogeneity is $\tau=0.302$, with 95\% credible interval
$[0.100,0.661]$. For a new true study effect, the posterior predictive
prevented fraction has median $0.386$ and 95\% predictive interval
$ 
[-0.285,\;0.732].
$ 

The posterior probability of any preventive effect in a new comparable study is
$$
\Pr(\PF_{+,\mathrm{new}}>0\mid\mathcal D)=0.926,
$$
while the probability of at least a $20\%$ preventive fraction is
$$
\Pr(\PF_{+,\mathrm{new}}>0.20\mid\mathcal D)=0.809.
$$

Table~\ref{tab:historical-models} summarizes the principal historical analyses and targeted sensitivity
analyses. The pooled center is notably stable. Across the four heterogeneity
priors in the contrast-based Bayesian model, the posterior median pooled
prevented fraction varies only from $0.380$ to $0.389$. In contrast, estimated
heterogeneity and the tails of the predictive distribution respond more visibly
to the heterogeneity prior.

\begin{table}[ht]
\centering
\caption{Historical synthesis under the principal frequentist and Bayesian
models. Bayesian entries are posterior medians; intervals are 95\% credible or
predictive intervals.}
\label{tab:historical-models}
\small
\begin{tabular}{p{4.0cm}cccc}
\toprule
Analysis &
Pooled $\PF_+$ &
$\tau$ &
Predictive $\PF_{+,\mathrm{new}}$ &
$\Pr(\PF_{+,\mathrm{new}}>0)$ \\
\midrule
REML--HK
& 0.397 $[0.180,0.556]$
& 0.284
& $[-0.265,0.712]$
& -- \\

Bayesian NNHM,
Half-N$(0,0.5^2)$
& 0.389 $[0.215,0.556]$
& 0.302
& 0.386 $[-0.285,0.732]$
& 0.926 \\

Bayesian NNHM,
Half-N$(0,0.25^2)$
& 0.382 $[0.233,0.529]$
& 0.253
& 0.379 $[-0.127,0.683]$
& 0.951 \\

Bayesian NNHM,
Half-$t_4(0,0.5)$
& 0.389 $[0.215,0.556]$
& 0.300
& 0.386 $[-0.285,0.732]$
& 0.927 \\

Bayesian NNHM,
DuMouchel prior
& 0.380 $[0.229,0.538]$
& 0.249
& 0.375 $[-0.170,0.700]$
& 0.947 \\

Bayesian arm-level,
normal random effects
& 0.433 $[0.222,0.599]$
& 0.385
& 0.430 $[-0.436,0.788]$
& 0.908 \\

Bayesian arm-level,
$t_4$ random effects
& 0.421 $[0.233,0.580]$
& 0.434
& 0.422 $[-0.584,0.794]$
& 0.909 \\
\bottomrule
\end{tabular}
\end{table}

The prior-sensitivity results are informative in two ways. First, reasonable
changes in the prior for $\tau$ have minimal effect on the historical pooled
center: the full range of posterior medians is less than one percentage point.
Second, the same changes have a noticeably larger effect on heterogeneity and
prediction. In this small meta-analysis, prior assumptions therefore matter
more for statements about transportability than for estimation of the
historical average treatment effect.

The aggregate arm-level analysis provides a different type of robustness check.
Avoiding the first-stage log-ratio approximation increases the posterior median
prevented fraction from $0.389$ to $0.433$, but does not change the substantive
conclusion. Under Gaussian random effects, the posterior median heterogeneity is
$\tau=0.385$ with 95\% credible interval $[0.171,0.768]$, and the predictive
prevented fraction has median $0.430$ with 95\% interval $[-0.436,0.788]$.
The posterior probabilities of any preventive effect and of at least a $20\%$
preventive fraction in a new comparable study are $0.908$ and $0.802$,
respectively.

Replacing Gaussian random effects by standardized $t_4$ random effects leaves
the pooled center similar at $0.421$, with 95\% credible interval
$[0.233,0.580]$, while increasing the posterior median heterogeneity to
$\tau=0.434$, with 95\% credible interval $[0.176,0.872]$. The corresponding
predictive prevented fraction has median $0.422$ and 95\% interval
$[-0.584,0.794]$. Thus the heavier-tailed model has a noticeably larger effect
on heterogeneity and prediction than on the historical pooled center.

The Stan arm-level analyses showed satisfactory computational behavior.
Across the Gaussian and standardized $t_4$ fits, rank-normalized
$\widehat R$ values were below $1.001$, effective sample sizes were large,
and there were no divergent transitions or maximum-tree-depth hits.
Posterior predictive checks showed no indication of gross arm-level model
misfit. Complete MCMC and posterior predictive diagnostics are reported in
the Supplementary Material.

Taken together, the principal models place the historical pooled preventive
fraction in a relatively narrow range, approximately $0.39$--$0.43$. The
uncertainty regarding a future setting is much less stable. This distinction is
important: the historical evidence is considerably more informative about the
average effect observed in those trials than about the treatment effect that
should be expected in a new population.

The effect-scale comparison reinforces the same point from a different
direction. The earlier direct prevented-fraction REML analysis gives a pooled
estimate of approximately $0.434$, whereas the preferred log-ratio REML
analysis gives $0.397$. The resulting difference of about $3.7$ percentage
points is substantially larger than the change produced by switching among the
reasonable heterogeneity priors in the Bayesian log-ratio model. For this
dataset, choosing an appropriate effect scale is therefore more consequential
for the pooled center than fine distinctions among defensible priors for
$\tau$.

A second-order bias-corrected log-response-ratio sensitivity analysis leads to
the same substantive conclusion. Using the Lajeunesse correction for both the
log response ratio and its sampling variance~\cite{Lajeunesse2015}, the REML pooled log ratio
changes from $-0.505$ to $-0.478$, corresponding to a prevented fraction of
$0.380$ with Hartung--Knapp 95\% CI $[0.183,0.529]$. The estimated
heterogeneity decreases from $0.284$ to $0.256$, and the 95\% prediction
interval becomes $[-0.210,0.682]$ on the prevented-fraction scale. The largest
study-level corrections occur for Koch and Tewari, as expected from their small
treatment means relative to within-arm variation. Full corrected study-level
results are reported in the Supplementary Material.

\subsection{Historical completeness: comparison with the Cochrane-13 evidence set}
\label{sec:cochrane13-brief}

The nine studies used for the principal analysis do not constitute the complete
historical permanent-tooth-surface evidence base. The 2013 Cochrane review
included 13 trials in the corresponding prevented-fraction synthesis, four of
which: Holm, Borutta, Hardman, and Gugwad, are not represented in the
nine-study log-ratio dataset~\cite{Marinho2013}. These additional trials cannot all be inserted
directly into the preferred log-ratio likelihood because of differences in
design, reconstruction, shared controls, and reported outcome structure. We
therefore retain the Cochrane-compatible prevented-fraction scale solely for
this historical-completeness analysis and do not combine it with the preferred
log-ratio synthesis.
The comparison is shown in Table~\ref{tab:cochrane-completeness}.

\begin{table}[ht]
\centering
\caption{Historical-completeness sensitivity analysis comparing the nine-study
subset with the complete 13-trial Cochrane permanent-surface evidence set.
All quantities in this table are calculated on the direct prevented-fraction
scale used for the historical-completeness analysis, rather than on the
preferred log-ratio scale. Because the direct Gaussian model is not constrained
to the natural range of the prevented fraction, predictive intervals may extend
outside that range.}
\label{tab:cochrane-completeness}
\small
\begin{tabular}{lcc}
\toprule
Quantity & Nine studies & Cochrane-13 \\
\midrule
REML pooled PF                 & 0.439 & 0.435 \\
REML heterogeneity SD          & 0.217 & 0.197 \\
Bayesian posterior median PF   & 0.436 & 0.433 \\
Bayesian posterior median $\tau$ & 0.238 & 0.214 \\
95\% posterior predictive interval
                               & $[-0.15,1.01]$ & $[-0.08,0.94]$ \\
\bottomrule
\end{tabular}
\end{table}

The historical center is essentially unchanged. Expanding the evidence set from
nine to thirteen trials changes the REML pooled prevented fraction from
$0.439$ to $0.435$, a difference of less than one percentage point, and the
Bayesian posterior medians are similarly close. If anything, the larger
historical set yields slightly less estimated heterogeneity and a somewhat
narrower predictive distribution.

The purpose of this comparison is not to define a second historical
meta-analysis, but to assess the robustness of the historical reference effect
to inclusion of the complete Cochrane permanent-surface evidence set. This
comparison is conducted on the direct prevented-fraction scale used for the
historical-completeness analysis and should therefore not be confused with the
preferred log-ratio analysis, for which the nine-study REML estimate is
$0.397$. On the common direct prevented-fraction scale, expanding the evidence
set from nine to 13 trials changes the pooled estimate only from $0.439$ to
$0.435$. The more detailed statistical analysis can therefore remain based on
the nine studies for which the preferred log-ratio and arm-level formulations
are available, while the Cochrane-13 analysis serves as a
historical-completeness sensitivity analysis. The study-specific design and
reconstruction issues for Holm, Borutta, Hardman, and Gugwad are documented in
the Supplementary Material.

\section{Contemporary Evidence and Evolution of the Preventive Effect}
\label{sec:contemporary}

The historical analyses establish a stable reference center near a $40\%$
prevented fraction. We now ask whether the incremental effect observed in
contemporary trials is of similar magnitude and whether it changes
systematically with calendar time. The continuous-outcome analysis uses the
14-study primary dataset described in Section~\ref{ssec:temporal}; binary  incidence outcomes 
are retained as a parallel evidence stream because they estimate a different
effect quantity.

\subsection{Continuous-outcome evidence and calendar-time evolution}
\label{sec:continuous-time}

Table~\ref{tab:modern-continuous} summarizes the contemporary studies considered
for the continuous-outcome analysis. Five studies enter the primary weighted
temporal model. Jayasinghe and Ghasemi remain sensitivity-only because their
sampling variances cannot be recovered uniquely from the published summaries.

\begin{table}[ht]
\centering
\caption{Contemporary continuous-outcome evidence. Crude prevented fractions
are $\PF_+=1-\exp(\LRR)$. ``Primary'' denotes inclusion in the 14-study
calendar-time analysis.}
\label{tab:modern-continuous}
\small
\begin{tabular}{p{2.7cm}c p{3.0cm}rr p{4.2cm}}
\toprule
Study & Year & Contrast & LRR & $\PF_+$ & Analysis status \\
\midrule
Mu\~noz-Mill\'an & 2018 & FV vs placebo
 & $-0.272$ & 0.238 & Primary \\
Latifi-Xhemajli & 2019 & FV vs usual care
 & $-0.664$ & 0.485 & Primary; endpoint-burden sensitivity \\
McMahon & 2020 & FV vs program control
 & $\phantom{-}0.000$ & 0.000 & Primary; endpoint-burden sensitivity \\
Wang & 2022 & FV vs no FV
 & $-0.473$ & 0.377 & Primary; approximately cluster-adjusted SE \\
Jayasinghe & 2022 & FV vs control
 & $-0.273$ & 0.239 & Variance sensitivity only \\
Ghasemi & 2025 & OHE+FV vs OHE
 & $-1.386$ & 0.750 & Variance sensitivity only \\
Zeng & 2025 & FV+OHE vs OHE
 & $\phantom{-}0.045$ & $-0.046$ & Primary \\
\bottomrule
\end{tabular}
\end{table}

\paragraph{Competing temporal models}
Table~\ref{tab:time-model-comparison} compares the candidate temporal specifications. The intercept-only
model has the smallest AICc, so the data do not require a calendar-time
moderator. Nevertheless, both the linear-time and
historical-versus-contemporary models lie within two AICc units of the
intercept-only model. Among models that allow the effect to vary with calendar
time, the simple linear specification is the most parsimonious; neither the
quadratic nor the three-era model gains sufficient fit to justify the
additional parameters.

\begin{table}[ht]
\centering
\caption{Maximum-likelihood comparison of alternative calendar-time models for
the 14-study primary continuous-outcome dataset. Akaike weights are descriptive,
not posterior model probabilities.}
\label{tab:time-model-comparison}
\small
\begin{tabular}{lrrr}
\toprule
Model & AICc & $\Delta$AICc & Akaike weight \\
\midrule
No calendar moderator              & 17.29 & 0.00 & 0.430 \\
Linear calendar time               & 18.13 & 0.84 & 0.283 \\
Historical vs contemporary era     & 18.73 & 1.44 & 0.209 \\
Quadratic calendar time            & 21.73 & 4.44 & 0.047 \\
Three calendar eras                & 22.58 & 5.29 & 0.031 \\
\bottomrule
\end{tabular}
\end{table}

\paragraph{Linear calendar-time model}
The REML--Hartung--Knapp fit gives
$$
\widehat\beta_0=-0.4768,
\qquad
\widehat\beta_1=0.0986
\quad\text{per decade},
$$
with residual heterogeneity
$ 
\widehat\tau=0.2787.
$ 
The 95\% Hartung--Knapp interval for the time coefficient is
$ 
[-0.0445,\;0.2417].
$ 
The point estimate therefore lies in the direction of attenuation, while the
interval remains compatible with no temporal trend.

The coefficient is directly interpretable on the ratio scale:
$
\exp(0.0986)=1.104.
$
Thus, under the fitted linear model, the treatment-to-control mean ratio is
multiplied by approximately $1.10$ for each later decade and, over the observed
calendar range, moves toward its null value of one. For $t=(\mathrm{Year}-2000)/10$,
$$
\widehat\theta(t)
=
-0.4768+0.0986t,
\qquad
\widehat{\PF}_+(t)
=
1-\exp\{\widehat\theta(t)\}.
$$

\begin{table}[ht]
\centering
\caption{Prevented fractions implied by the 14-study linear calendar-time
meta-regression. These are fitted cross-study values, not repeated estimates
for one fixed population.}
\label{tab:time-pf}
\small
\begin{tabular}{ccc}
\toprule
Calendar year & Fitted log ratio & Fitted $\PF_+$ \\
\midrule
2000 & $-0.4768$ & 0.379 \\
2005 & $-0.4275$ & 0.348 \\
2010 & $-0.3782$ & 0.315 \\
2020 & $-0.2796$ & 0.244 \\
2025 & $-0.2303$ & 0.206 \\
\bottomrule
\end{tabular}
\end{table}

The twenty-year comparison is particularly transparent: the fitted prevented
fraction decreases from approximately $34.8\%$ in 2005 to $20.6\%$ in 2025,
a difference of about 14 percentage points. Relative to the 2005 fitted value,
this is an attenuation of roughly $41\%$. These quantities summarize a
study-level temporal trajectory and should not be interpreted as the effect
that the same fixed population would have exhibited at two calendar dates.

\begin{figure}[ht]
\centering
\includegraphics[width=0.90\textwidth]{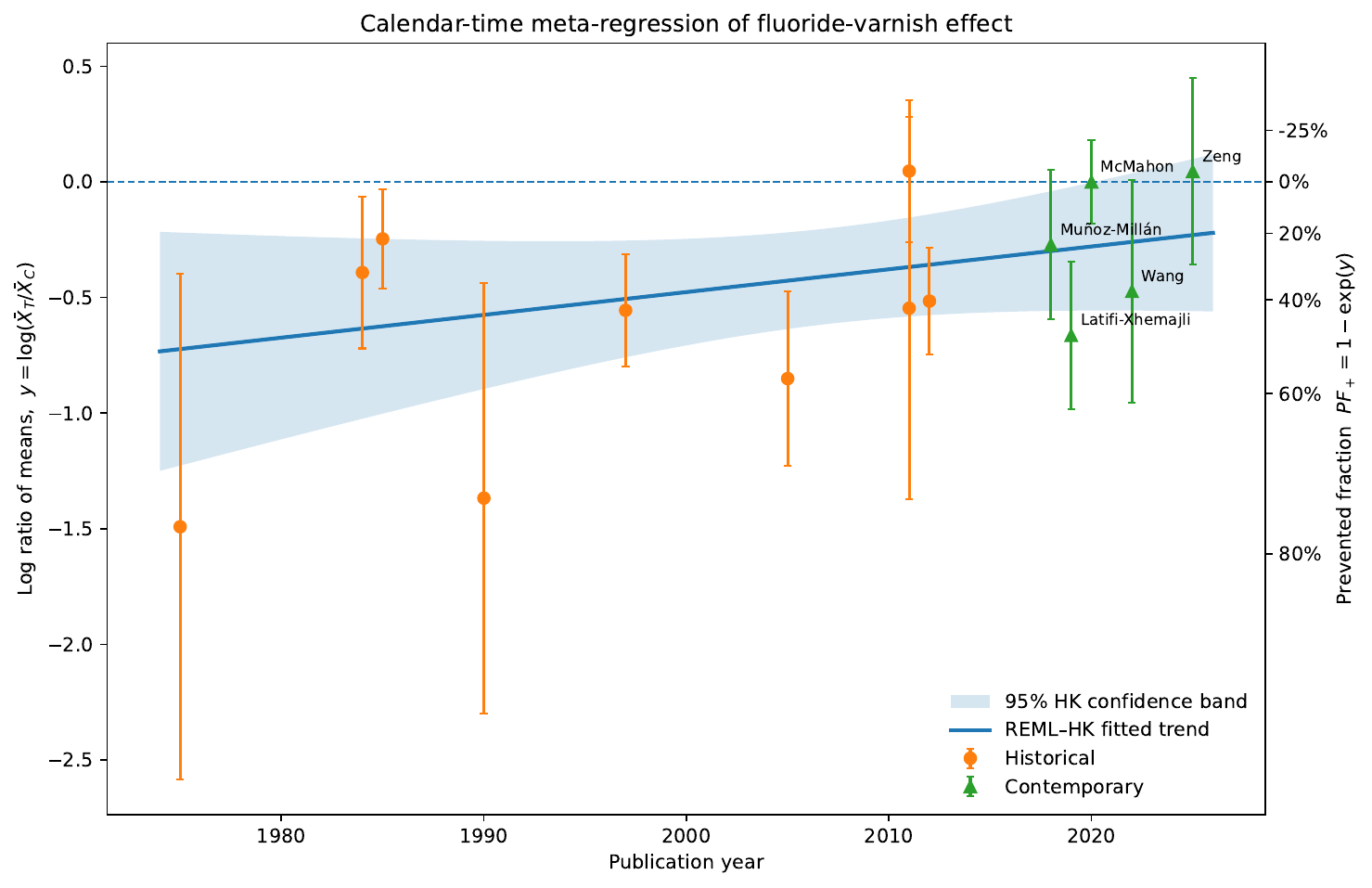}
\caption{Study-specific log ratios of treatment-to-control mean caries outcomes
plotted against calendar year for the 14-study primary dataset. Historical and
contemporary studies are distinguished by plotting symbols. Vertical segments
show study-specific 95\% sampling intervals. The solid curve is the fitted
linear random-effects meta-regression and the band is the corresponding 95\%
Hartung--Knapp uncertainty interval for the fitted mean. The right-hand axis
gives the equivalent prevented fraction $\PF_+=1-\exp(y)$. Wang is included
using the approximately cluster-adjusted sampling variance based on the
published intraclass correlation coefficient and planning cluster size.}
\label{fig:calendar-time}
\end{figure}

Figure~\ref{fig:calendar-time} makes both features of the result visible:
considerable between-study variation remains, but the fitted trajectory moves
upward toward a log ratio of zero.

\paragraph{Era comparison and Bayesian analysis}
A coarser historical-versus-contemporary comparison points in the same
direction without imposing a linear calendar trend. The historical nine-study
log-ratio analysis gives a pooled prevented fraction of $39.7\%$. Separately,
pooling the five contemporary studies in the primary 14-study dataset gives a
REML center of approximately $22.7\%$ (Hartung--Knapp 95\% interval
approximately $-12.9\%$ to $47.1\%$). In the corresponding era
meta-regression, the contemporary-minus-historical coefficient is
$
\widehat{\beta}_E=0.242,
$
with 95\% Hartung--Knapp interval
$
[-0.190,\,0.675].
$
Thus the fitted centers differ appreciably, but the era contrast is imprecisely
estimated.

The Bayesian calendar-time model gives a posterior median
$$
\operatorname{median}(\beta_1\mid\data)=0.091
$$
per decade, with 95\% credible interval
$ 
[-0.036,\;0.231],
$  
and
$ 
\Pr(\beta_1>0\mid\data)=0.924.
$ 
The Bayesian and frequentist analyses therefore agree on the direction of the
point estimate while retaining substantial uncertainty about its magnitude.

The direction is not driven by a single contemporary study. Deleting
Mu\~noz-Mill\'an, Latifi-Xhemajli, McMahon, Zeng, or Wang one at a time gives
calendar slopes between approximately $0.076$ and $0.116$ log-ratio units per
decade; every point estimate remains positive. The variance-audit sensitivities
give the same directional result. Adding Jayasinghe under a representative
quasi-Poisson overdispersion assumption gives a slope near $0.094$, and then
adding Ghasemi under a representative cluster ICC of $0.05$ gives a slope near
$0.085$. Across the full joint Jayasinghe/Ghasemi variance grid, the fitted
slope ranges from approximately $0.071$ to $0.091$. Details are provided in the
Supplementary Material.

\subsection{Contemporary evidence from binary caries outcomes}
\label{sec:binary-modern}

Several contemporary trials report child-level caries incidence rather than a
continuous DMF-type outcome. These outcomes estimate a different quantity and
should therefore not be combined with the log-ratio-of-means analysis. They do,
however, provide a useful parallel assessment of whether a modest contemporary
varnish contrast is also evident on a binary incidence scale.

For this evidence stream we use the log risk ratio
$$
r_i=\log\left(\frac{p_{Ti}}{p_{Ci}}\right),
$$
with $r_i<0$ favoring a varnish-containing strategy. Five contemporary
comparisons---Mu\~noz-Mill\'an, McMahon, Wang, Zeng, and He---give crude risk
ratios of approximately $0.81$, $0.85$, $0.72$, $1.08$, and $0.81$,
respectively~\cite{Munoz2018,McMahon2020,Wang2022,Zeng2025,He2026}.

An exploratory REML random-effects synthesis gives
$$
\widehat\rho=-0.225,
\qquad
\widehat\tau_R=0.073,
\qquad
\exp(\widehat\rho)\approx0.80,
$$
with an approximate 95\% interval for the pooled risk ratio of
$$
[0.71,\;0.89].
$$
Under a Half-Normal$(0,0.5^2)$ heterogeneity prior, the posterior median risk
ratio is also approximately $0.80$, with a 95\% credible interval near
$[0.69,1.00]$ and a substantially wider predictive interval near
$[0.56,1.24]$.

The numerical agreement with the contemporary continuous-outcome center
(roughly a 20--23\% relative reduction) is useful, but it should not be
overstated. The two evidence streams use different estimands, and several
studies contribute to both through different outcomes. The binary analysis is
therefore treated as a complementary contemporary outcome analysis rather than
as an independent replication or as part of a common pooled effect.

\section{Why Might the Incremental Preventive Effect Be Declining?}
\label{sec:why-decline}

The temporal analysis describes a secular association; it does not identify a
mechanism. A plausible interpretation is that the preventive counterfactual has
changed rather than that fluoride varnish has become pharmacologically weaker.
Contemporary trial populations are more likely to be exposed to fluoride
toothpaste and may also receive fluoridated water, supervised brushing,
oral-health education, preventive dental care, fissure sealants, or structured
community prevention programs. A contemporary trial may therefore ask how much
additional benefit varnish provides after other preventive measures are already
operating.

This changing counterfactual can reduce the absolute treatment contrast even
without changing the relative effect. If $m_C$ is the expected control mean
and $\theta$ the log treatment-to-control mean ratio, then
$$
m_T=m_C\exp(\theta),
\qquad
\Delta(m_C,\theta)
=
m_C\{1-\exp(\theta)\}
=
m_C\PF_+.
$$
Thus, lower control-group disease burden necessarily reduces the absolute
amount of disease available to prevent when $\PF_+$ is fixed. The relative
effect may also become smaller if background preventive measures
disproportionately remove the component of disease most responsive to fluoride
varnish, leaving residual disease concentrated in higher-risk or less
preventable children or surfaces.

Trial design and reporting also differ across calendar time. The contemporary
studies in the present analysis include active or structured control conditions,
cluster-randomized designs, and more explicit attention to design-adjusted
uncertainty. Calendar time therefore combines changes in background prevention,
population risk, outcome measurement, and trial design. The calendar
coefficient should consequently be interpreted as a summary of secular change
rather than as a causal parameter.

\subsection{Explanatory moderator analysis}
\label{sec:explanatory}

To examine whether measurable features of the preventive environment account
for part of the time signal, we fitted a prespecified Bayesian moderator
sequence to the same 14-study primary continuous-outcome dataset. Let
$B_i=1$ when the control arm received a protocol-specified organized
non-varnish preventive intervention in addition to ordinary care, and $B_i=0$
otherwise. The model
$$
\theta_i
\sim
N\left(
\beta_0+\beta_1t_i+\beta_2B_i,\tau^2
\right)
$$
uses the same weakly regularizing priors as the calendar-time analysis.

The unadjusted 14-study posterior median time slope is $0.091$ per decade
(95\% CrI $-0.036$ to $0.231$), with
$\Pr(\beta_1>0\mid\data)=0.924$. After adding organized background prevention,
the median time slope decreases to
$ 
0.057
\quad
[-0.071,\;0.213],
$ 
and
$ 
\Pr(\beta_1>0\mid\data)=0.804.
$ 
The background-prevention coefficient has posterior median
$ 
\beta_2=0.178
\quad
[-0.195,\;0.505],
$ 
with $\Pr(\beta_2>0\mid\data)=0.834$. Thus, adjustment for organized background prevention reduces the posterior
median calendar slope by approximately $37\%$, although the moderator
coefficient itself remains imprecisely estimated.

Because observed control-arm burden appears in the denominator of the log ratio,
regressing the observed log ratio naively on the observed control mean can
induce a mathematical association~\cite{Schmid1998,Arends2000}. We therefore
also use an error-aware arm formulation in which the latent control mean enters
the treatment-effect model. With an additional standardized log control-burden
term, the posterior median calendar coefficient is $0.053$ (95\% CrI
$-0.074$ to $0.212$), with posterior probability $0.789$ of being positive.
The background-prevention coefficient remains positive at $0.171$ on the
posterior median, while the burden coefficient is $-0.060$ (95\% CrI
approximately $-0.240$ to $0.135$). The burden term therefore adds little
additional explanation beyond the organized-prevention indicator. Adding the burden term therefore changes the fitted calendar coefficient little
further beyond the adjustment for organized prevention.

These moderator analyses sharpen the interpretation but do not establish
mediation. Calendar time and preventive intensity are correlated, the number
of studies is small, and control-arm burden is not measured on a common
clinical scale across teeth, surfaces, endpoint burdens, and increments.
The most defensible conclusion is therefore comparative: adjustment for
measurable background prevention is associated with a non-negligible reduction
in the fitted calendar coefficient, but the available study-level data cannot
distinguish the underlying secular mechanisms precisely.

This interpretation also resolves an apparent paradox. A declining incremental
effect of varnish can coexist with improving population oral health. If the
preventive environment improves outcomes in both randomized groups, a smaller
additional contrast for varnish can be a consequence of successful background
prevention rather than evidence of treatment failure. Similar secular
improvements in control-group outcomes have been documented in other clinical
areas as standards of care have improved~\cite{Hong2011,Garofalo2024}.

\section{Discussion}
\label{sec:discussion}

This analysis has two principal findings. First, the historical
fluoride-varnish effect is stable at the level of the pooled center. The
preferred nine-study REML log-ratio analysis gives a prevented fraction of
$39.7\%$; Bayesian contrast, aggregate arm-level, and heavy-tailed
random-effects models place the historical center in approximately the same
$0.39$--$0.43$ range. Expanding from the nine-study core to the 13-trial
Cochrane permanent-surface set changes the pooled prevented fraction by less
than one percentage point. The historical reference is therefore robust to
reasonable changes in the heterogeneity prior, random-effects distribution,
first-stage log-ratio approximation, and historical study set.

The historical analysis also illustrates why a stable pooled mean should not be
confused with precise transportability. Prediction intervals are much wider
than confidence or credible intervals for the average effect, and they are more
sensitive to the heterogeneity prior and random-effects distribution. In a
small meta-analysis, this is not a technical footnote: uncertainty about
$\tau$ directly limits how confidently an historical average can be exported
to a new population.

Effect-scale choice matters as well. Modeling a ratio-derived effect on the
log-ratio scale moves the pooled historical estimate from approximately $0.434$
under the direct Gaussian prevented-fraction sensitivity analysis to $0.397$
under the preferred REML log-ratio analysis. By comparison, switching among
reasonable Bayesian heterogeneity priors changes the pooled center by less than
one percentage point. The completed second-order response-ratio sensitivity
changes the preferred estimate only modestly, from $39.7\%$ to $38.0\%$.
Together with the aggregate arm-level analysis, this indicates that the main
historical conclusion is not driven by the first-order response-ratio
approximation~\cite{Friedrich2011,Bakbergenuly2020,Lajeunesse2015}.

Second, the contemporary extension is compatible with attenuation of the
incremental treatment contrast. In the 14-study primary temporal model, the
REML calendar coefficient is $0.0986$ log-ratio units per decade and the
Bayesian posterior median is $0.091$. Both are positive, meaning that the
treatment-to-control ratio moves toward one in later studies. The fitted
prevented fraction falls from approximately $34.8\%$ in 2005 to $20.6\%$ in
2025. Separately, pooling the five primary contemporary studies gives a
prevented-fraction center of approximately $22.7\%$.

The evidence should nevertheless be described as suggestive rather than
conclusive. The Hartung--Knapp interval for the calendar slope includes zero,
and the intercept-only model has the smallest AICc. The linear-time and
historical-versus-contemporary specifications are close competitors, however,
and all leave-one-contemporary-study-out slopes remain positive. Adding
Jayasinghe and Ghasemi over broad, explicitly stated variance assumptions also
preserves a positive point estimate. Thus the data show a consistent directional pattern toward attenuation, but
they do not statistically require a temporal moderator.

The explanatory analysis gives the temporal pattern a more substantive
interpretation. Adjustment for protocol-specified organized prevention in the
control condition reduces the Bayesian median time slope from $0.091$ to $0.057$
per decade. Adding an error-aware proxy for control-group caries burden changes
it only slightly further, to $0.053$. These estimates are too imprecise to
establish mediation, but they are consistent with the hypothesis that part of
the apparent secular attenuation reflects a stronger preventive counterfactual
and does not require a weakening varnish mechanism.

The binary incidence analysis provides a complementary contemporary effect
estimate, with an exploratory pooled risk ratio of approximately $0.80$.
This should be viewed as a parallel outcome-based sensitivity check, not a
second independent dataset, because several trials contribute both continuous
and binary outcomes and the estimands differ. Its value is that a modest
contemporary varnish contrast is not confined to one particular continuous
effect definition.

Several limitations remain. The contemporary extension is not a comprehensive
updated systematic review. Dentition, endpoint definition, follow-up,
comparator, and randomization unit differ across studies, and Wang requires an
approximate cluster adjustment rather than a published cluster-robust variance.
Jayasinghe and Ghasemi remain sensitivity-only studies because the published
summaries do not identify unique design-adjusted sampling variances. The analyses
are based on aggregate data and cannot recover individual risk factors,
zero-inflation, or within-child dependence. Finally, calendar time remains
confounded with changes in prevention, population risk, trial conduct, and
reporting. These limitations argue for restrained interpretation and against a
larger multivariable meta-regression with the present evidence base.

The combined evidence therefore changes the framing more than it changes the
historical result. The older trials support a substantial average preventive
effect. The more contemporary question is how much additional protection
varnish provides when introduced into populations already receiving substantial
preventive care. Historical and contemporary treatment contrasts need not be
numerically identical because they do not represent the same preventive
counterfactual.

\section{Conclusions}
\label{sec:conclusions}

The historical fluoride-varnish trials support a substantial average preventive
effect. Across the preferred frequentist and Bayesian analyses, the historical
pooled center is close to $40\%$, and expansion to the complete 13-trial
Cochrane permanent-surface set leaves that center essentially unchanged.
Uncertainty about prediction to a new setting is materially larger than
uncertainty about the historical average.

The 14-study contemporary extension does not establish a temporal decline, but
it is consistently compatible with attenuation. The fitted log
treatment-to-control ratio moves toward zero with calendar time, the
contemporary pooled center is smaller than the historical center, and the
direction persists across leave-one-study and variance-sensitivity analyses.
At the same time, the no-moderator model has the lowest AICc and the
calendar-time interval includes zero.

A plausible interpretation, consistent with these analyses, is therefore not
that fluoride varnish has lost biological efficacy, but that its incremental
contribution may be smaller when added to a stronger preventive environment.
The moderator analysis is consistent with this explanation: explicit adjustment
for organized background prevention reduces the fitted calendar slope
substantially, although the available study-level evidence remains too sparse
for a causal decomposition. A future comprehensive evidence update should
prioritize harmonized contemporary endpoints, design-adjusted variances, and
explicit characterization of background preventive care in control groups,
rather than treating calendar year itself as the mechanism.

\medskip

\paragraph{Supplementary Material and Reproducibility}

Supplementary material and the complete computational code supporting this
article are available at the accompanying GitHub repository:
\url{github.com/BraniV/FluorideVarnish}.

The repository contains the Supplementary Material, together with the R and
Stan code used to reproduce the analyses reported in the paper. The
Supplement provides the detailed frequentist REML--Hartung--Knapp analysis of
the nine historical trials; the historical-completeness analysis using the
13-trial Cochrane evidence set; Bayesian normal--normal hierarchical analyses
and sensitivity to the heterogeneity prior; aggregate arm-level Gaussian and
heavy-tailed models fitted in Stan, including convergence and posterior
predictive diagnostics; the contemporary continuous-outcome extraction,
variance audit, temporal-model comparisons, influence analyses, and
variance-sensitivity calculations; the supporting analysis of contemporary
binary caries outcomes; and the explanatory meta-regressions involving
organized background prevention and control-arm caries burden. The
Supplement also records the software versions, random-number seeds, and other
computational details needed to reproduce the reported results.

\paragraph{Funding}
This work was supported in part by the National Science Foundation under
Grant No.~2515246. Partial support for Sara Antonijevic was provided by the
Thomas W. Powell '62 Endowed Graduate Fellowship at Texas A\&M University.

\paragraph{Acknowledgments}
Brani Vidakovic thanks Susan Griffin and Laurie Barker for their collaboration
on the meta-analytic work involving the historical data during his
Intergovernmental Personnel Act assignment with the Dental Group at the
Centers for Disease Control and Prevention in Atlanta, Georgia.
The findings and conclusions in this article are those of the authors and do
not necessarily represent the official position of the Centers for Disease
Control and Prevention.

\end{document}